\documentclass[aoas,MSNbibl,nameyear,rotating,dvips]{arximspdf}
\usepackage{multirow,dcolumn}
\usepackage{graphicx}

\doi{10.1214/12-AOAS593} 
\volume{7}
\issue{1}
\pubyear{2013}
\firstpage{443}
\lastpage{470}

\makeatletter
\newcolumntype{d}[1]{D{.}{.}{#1}}
\newcommand{\rrvert}{\vert}
\newcommand{\llvert}{\vert}

\newcommand{\sgn}{\operatorname{sgn}}

\newcommand\cA{\mathcal{A}}
\newcommand\E{\mathbb{E}}
\newcommand\cN{\mathcal{N}}
\newcommand\cP{\mathcal{P}}
\newcommand\cT{\mathcal{T}}
\newcommand\cX{\mathcal{X}}
\newcommand\cY{\mathcal{Y}}

\newcommand{\argmax}{\mathop{\arg\max}}
\newcommand{\argmin}{\mathop{\arg\min}}

\newcommand\spacingset[1]{}

\setattribute{abstract}{width}{290pt}
\makeatother

\begin{document}
\begin{frontmatter}

\title{Estimating treatment effect heterogeneity in randomized program
evaluation\thanksref{T1}}
\thankstext{T1}{Supported by NSF Grant SES--0752050.
The proposed methods can be implemented via
open-source software \texttt{FindIt} [Ratkovic and Imai (\citeyear{RaIm12})], which is
freely available at the Comprehensive R Archive Network (\protect\url{http://cran.r-project.org/package=FindIt}).  This software also
contains the results of our empirical analysis.}
\runtitle{Estimating treatment effect heterogeneity}

\begin{aug}
\author[A]{\fnms{Kosuke} \snm{Imai}\corref{}\ead[label=e1]{kimai@princeton.edu}\ead[label=u1,url]{http://imai.princeton.edu}}
\and
\author[A]{\fnms{Marc} \snm{Ratkovic}\ead[label=e2]{ratkovic@princeton.edu}}
\runauthor{K. Imai and M. Ratkovic}
\affiliation{Princeton University}
\address[A]{Department of Politics\\
Princeton University\\
Princeton, New Jersey 08544\\
USA\\
\printead{e1}\\
\phantom{E-mail:\ }\printead*{e2}\\
\printead{u1}}
\end{aug}

\received{\smonth{1} \syear{2012}}
\revised{\smonth{9} \syear{2012}}

%
\begin{abstract}
When evaluating the efficacy of social programs and medical
treatments using randomized experiments, the estimated overall
average causal effect alone is often of limited value and the
researchers must investigate when the treatments do and do not work.
Indeed, the estimation of treatment effect heterogeneity plays an
essential role in (1) selecting the most effective treatment from a
large number of available treatments, (2) ascertaining
subpopulations for which a treatment is effective or harmful, (3)
designing individualized optimal treatment regimes, (4) testing for
the existence or lack of heterogeneous treatment effects, and (5)
generalizing causal effect estimates obtained from an experimental
sample to a target population. In this paper, we formulate the
estimation of heterogeneous treatment effects as a variable
selection problem. We propose a method that adapts the Support
Vector Machine classifier by placing separate sparsity constraints
over the pre-treatment parameters and causal heterogeneity
parameters of interest. The proposed method is motivated by and
applied to two well-known randomized evaluation studies in the
social sciences. Our method selects the most effective voter
mobilization strategies from a large number of alternative
strategies, and it also identifies the characteristics of workers
who greatly benefit from (or are negatively affected by) a job
training program. In our simulation studies, we find that the
proposed method often outperforms some commonly used alternatives.
\end{abstract}

%

\end{frontmatter}

\section{Introduction and motivating applications}
\label{secintro}

While the average treatment effect can be easily estimated without
bias in randomized experiments, treatment effect heterogeneity
plays an essential role in evaluating the efficacy of social programs
and medical treatments. We define treatment effect heterogeneity as
the degree to which different treatments have differential causal
effects on each unit. For example, ascertaining subpopulations for
which a treatment is most beneficial (or harmful) is an important goal
of many clinical trials. However, the most commonly used method,
subgroup analysis, is often inappropriate and remains one of the most
debated practices in the medical research community
[e.g., \citet{roth05,laga06}]. Estimation of treatment effect
heterogeneity is also important when (1) selecting the most effective
treatment among a large number of available treatments, (2) designing
optimal treatment regimes for each individual or a group of
individuals
[e.g., \citet
{mans04,pineetal07,moodplatkram09,imaistra11,cai2011,guntzhumurp11,Qian2011}],
(3) testing the existence or
lack of heterogeneous treatment effects
[e.g., \citet{gailsimo85,davi92,crumetal08}], and (4)
generalizing causal effect estimates obtained from an experimental
sample to a target population [e.g., \citet
{fran09,colestua10,hartetal10,greekern10a,stuaetal11}]. In all of these
cases, the
researchers must infer how treatment effects vary across individual
units and/or how causal effects differ across various treatments.

Two well-known randomized evaluation studies in the social sciences
serve as the motivating applications of this paper. Earlier analyses
of these data sets focused upon the estimation of the overall average
treatment effects and did not systematically explore treatment effect
heterogeneity. First, we analyze the get-out-the-vote (GOTV) field
experiment where many different mobilization techniques were randomly
administered to registered New Haven voters in the 1998 election
[\citet{gerbgree00}]. The original experiment used an incomplete,
unbalanced factorial design, with the following four factors: a
personal visit, 7 possible phone messages, 0 to 3 mailings, and one of
three appeals applied to visit and mailings (civic duty, neighborhood
solidarity, or a close election). The voters in the control group did
not receive any of these GOTV messages. Additional information on
each voter includes age, residence ward, whether registered for a
majority party, and whether the voter abstained or did not vote in the 1996 election. Here, our
goal is to identify a set of GOTV mobilization strategies that can
best increase turnout. Given the design, there exist 193 unique
treatment combinations, and the number of observations assigned to
each treatment combination ranges dramatically, from the minimum of 4
observations (visited in person, neighbor/civic-neighbor phone appeal,
two mailings, with a civic appeal) to the maximum of $2956$ (being
visited in person, with any appeal). The methodological challenge is
to extract useful information from such sparse data.

The second application is the evaluation of the national supported
work (NSW) program, which was conducted from 1975 to 1978 over 15
sites in the United States.\vadjust{\goodbreak} Disadvantaged workers who qualified for
this job training program consisted of welfare recipients, ex-addicts,
young school dropouts, and ex-offenders. We consider the binary
outcome indicating whether the earnings increased after the job
training program (measured in 1978) compared to the earnings before
the program (measured in 1975). The pre-treatment covariates include
the 1975 earnings, age, years of education, race, marriage status,
whether a worker has a college degree, and whether the worker was
unemployed before the program (measured in 1975). Our analysis
considers two aspects of treatment effect heterogeneity. First, we
seek to identify the groups of workers for whom the training program
is beneficial. The program was administered to the heterogeneous
group of workers and, hence, it is of interest to investigate whether
the treatment effect varies as a function of individual
characteristics. Second, we show how to generalize the results based
on this experiment to a target population. Such an analysis is
important for policy makers who wish to use experimental results to
decide whether and how to implement this program in a target
population.

To address these methodological challenges, we formulate the
estimation of heterogeneous treatment effects as a variable selection
problem [see also \citet{guntzhumurp11,imaistra11}]. We propose
the Squared Loss Support Vector Machine (L2-SVM) with separate LASSO
constraints over the pre-treatment and causal heterogeneity parameters
(Section~\ref{secmodel}). The use of two separate constraints
ensures that variable selection is performed separately for variables
representing alternative treatments (in the case of the GOTV
experiment) and/or treatment-covariate interactions (in the case of
the job training experiment). Not only do these variables differ
qualitatively from others, they often have relatively weak predictive
power. The proposed model avoids the ad-hoc variable selection of
existing procedures by achieving optimal classification and variable
selection in a single step
[e.g., \citet{guntzhumurp11,imaistra11}]. The model also
directly incorporates sampling weights into the estimation procedure,
which are useful when generalizing the causal effects estimates
obtained from an experimental sample to a target
population.

To fit the proposed model with multiple regularization constraints, we
develop an estimation algorithm based on a generalized
cross-validation (GCV) statistic. When the derivation of an optimal
treatment regime rather than the description of treatment effect
heterogeneity is of interest, we can replace the GCV statistic with
the average effect size of the optimal treatment rule
[\citet{imaistra11,Qian2011}]. The proposed methodology with the GCV
statistic does not require cross-validation and hence is more
computationally efficient than the commonly used methods for
estimation of treatment effect heterogeneity such as Boosting
[\citet{Freund1999,LeBlanc2010}], Bayesian additive regression trees
(BART) [\citet{chipgeormccu10,greekern10}], and other tree-based
approaches [e.g., \citet{Su2009,imaistra11,Lipkovich2011},\vadjust{\goodbreak} \citet{Loh2012,Kang2012}]. While most similar
to a Bayesian logistic
regression with noninformative prior [\citet{Gelman2008}], the proposed
method uses LASSO constraints to produce a parsimonious model.

To evaluate the empirical performance of the proposed method, we
analyze the aforementioned two randomized evaluation studies
(Section~\ref{secapplications}). We find that personal visits are
uniformly more effective than any other treatment method, while
sending three mailings with a civic duty message is the most effective
treatment without a visit. In addition, every mobilization strategy
with a phone call, but no personal visit, is estimated to have either a negative or negligible positive effect.
For the job training study, we find that the program
is most effective for low-education, high income Non-Hispanics,
unemployed blacks with some college, and unemployed Hispanics with
some high school. In contrast, the program would be least effective
when administered to old, unemployed recipients, unmarried whites with a high
school degree but no college, and high earning Hispanics with no
college.

Finally, we conduct simulation studies to compare the performance of
the proposed methodology with that of various alternative methods
(Section~\ref{secsimulations}). The proposed method admits the
possibility of no treatment effect and yields a low false discovery
rate, when compared to the nonsparse alternative methods that
\emph{always} estimate some effects. Despite reductions in false
discovery, the method remains statistically powerful. We find that
the proposed method has a comparable discovery rate and competitive
predictive properties to these commonly used alternatives.


\section{The proposed methodology}
\label{secmodel}

In this section we describe the proposed methodology by presenting
the model and developing a computationally efficient estimation
algorithm to fit the model.

\subsection{The framework}

We describe our method within the potential outcomes framework of
causal inference. Consider a simple random sample of $N$ units from
population $\cP$, with a possibly different target population of
inference $\cP^\ast$. For example, the researchers and policy makers
may wish to apply the GOTV mobilization strategies and the job
training program to a population, of which the study sample is not
representative. We consider a multi-valued treatment variable $T_i$,
which takes one of $(K+1)$ values from $\cT\equiv\{0, 1, \ldots, K\}$
where $T_i=0$ means that unit $i$ is assigned to the control
condition. In the GOTV study, we have a total of 193 treatment
combinations ($K=193$), whereas the job training program corresponds to
a binary treatment variable ($K=1$). The potential outcome under
treatment $T_i=t$ is denoted by $Y_i(t)$, which has support $\cY$.
Thus, the observed outcome is given by $Y_i = Y_i(T_i)$ and we define
the causal effect of treatment $t$ for unit~$i$ as
$Y_i(t)-Y_i(0)$.

Throughout, we assume that there is no interference among units, there
is a unique version of each treatment,\vadjust{\goodbreak} each unit has nonzero
probability of assignment to each treatment level, and the treatment
level is independent of the potential outcomes, possibly conditional
on observed covariates [\citet{rubi90,roserubi83}]. Such
assumptions are met in randomized experiments, which are the focus of
this paper. Under these assumptions, we can identify the average
treatment effect (ATE) for each treatment $t$, $\tau(t) =
\E(Y_i(t) - Y_i(0))$. In observational studies, additional
difficulty arises due to the possible existence of unmeasured
confounders.

One commonly encountered problem related to treatment effect
heterogeneity requires selecting the most effective treatment from a
large number of alternatives using the causal effect estimates from a
finite sample. That is, we wish to identify the treatment condition
$t$ such that $\tau(t)$ is the largest, that is, $t =
\argmax_{t^\prime\in\cT} \tau(t^\prime)$. We may also be
interested in identifying a subset of the treatments whose ATEs are
positive. When the number of treatments $K$ is large as in the GOTV
study, a~simple strategy of subsetting the data and conducting a
separate analysis for each treatment suffers from the lack of power
and multiple testing problems.

Another common challenge addressed in this paper is identifying groups
of units for which a treatment is most beneficial (or most harmful), as
in the job training program study. Often, the number of available
pre-treatment covariates, $X_i \in\cX$, is large, but the
heterogeneous treatment effects can be characterized parsimoniously
using a subset of these covariates, $\widetilde{X}_i \in
\widetilde{\cX} \subset\cX$. This problem can be understood as
identifying a sparse representation of the conditional average
treatment effect (CATE), using only a subset of the covariates. We
denote the CATE for a unit with covariate profile $\tilde{x}$ as
$\tau(t; \tilde x) = \E(Y_i(t) - Y_i(0) \mid\widetilde{X}_i =
\tilde{x})$, which can be estimated as the difference in predicted
values under $T_i=t$ and $T_i=0$ with $\widetilde{X}_i = \tilde x$.
The sparsity in covariates greatly eases interpretation of this model.

We next turn to the description of the proposed model that combines
optimal classification and variable selection to estimate treatment
effect heterogeneity. For the remainder of the paper, we focus on the
case of binary outcomes, that is, $\cY= \{0, 1\}$. However, the
proposed model and algorithm can be extended easily to nonbinary
outcomes by modifying the loss function. We choose to model binary
outcomes with the L2-SVM to illustrate our proposed methodology
because it presents one of the most difficult cases for implementing
two separate LASSO constraints. As we discuss below, our method can be
simplified when the outcome is nonbinary (e.g., continuous, counts,
multinomial, hazard) or the causal estimand of interest is
characterized on a log-odds scale (with a logistic loss). In
particular, readily available software can be adapted to handle these
cases [\citet{Friedman2010}].

\subsection{The model}

In modeling treatment effect heterogeneity, we transform the observed
binary outcome to $Y_i^\ast=2Y_i-1 \in\{\pm1\}$. We then relate the
estimated\vadjust{\goodbreak} outcome $\widehat{Y}_i \in\{\pm1 \}$ and the estimated
latent variable $\widehat{W}_i \in\Re$, as
\[
\widehat{Y}_i  = \sgn{( \widehat{W}_i) }\qquad
\mbox{where } \widehat{W}_i = \hat\mu+ \hat\beta^\top
Z_i + \hat\gamma^\top V_i,
\]
$Z_i$ is an $L_Z$ dimensional vector of treatment effect heterogeneity
variables, and $V_i$ is an $L_V$ dimensional vector containing the
remaining covariates. For example, when identifying the most
efficacious treatment condition among many alternative treatments,
$Z_i$ would consist of $K$ indicator variables (e.g., different
combinations of mobilization strategies), each of which is
representing a different treatment condition. In contrast, $V_i$ would
include pre-treatment variables to be adjusted (e.g., age, party
registration, turnout history). Similarly, when identifying groups of
units most helped (or harmed) by a treatment, $Z_i$ would include
variables representing interactions between the treatment variable
(e.g., the job training program) and the pre-treatment covariates of
interest (e.g., age, education, race, prior employment status and
earnings). In this case, $V_i$ would include all the main effects of
the pre-treatment covariates. Thus, we separate the causal
heterogeneity variables of interest from the rest of the variables.
We do not impose any restriction between main and interaction effects
because some covariates may not predict the baseline outcome but do
predict treatment effect heterogeneity. Finally, we choose the linear
model because it allows for easy interpretation of interaction terms.
However, the researchers may also use the logistic or other link
function within our framework.

In estimating $(\beta, \gamma)$, we adapt the support vector machine
(SVM) classifier and place separate LASSO constraints over each set of
coefficients [\citet{Vapnik1995,Tibshirani1996,Bradley1998,Zhang2006}].
Our model differs from the standard model by allowing
$\beta$ and $\gamma$ to have separate LASSO constraints. The model is
motivated by the qualitative difference between the two parameters,
and also by the fact that often causal heterogeneity variables have
weaker predictive power than other variables. Specifically, we
formulate the SVM as a penalized squared hinge-loss objective function
(hereafter L2-SVM) where the hinge-loss is defined as $|x|_+ \equiv
\max(x,0)$ [\citet{Wahba2002}]. We focus on the L2-SVM, rather than the
L1-SVM, because it returns the standard difference-in-means estimate
for the treatment effect in the absence of pre-treatment
covariates.

With two separate $l_1$ constraints to
generate sparsity in the covariates, our estimates are given by
\[
(\hat\beta, \hat\gamma) = \argmin_{(\beta,\gamma)} \sum
_{i=1}^n w_i\cdot\bigl\llvert
1-Y_i^\ast \cdot \bigl(\mu+ \beta^\top
Z_{i} + \gamma^\top V_i \bigr)\bigr
\rrvert_+^2 +\lambda_Z \sum_{j=1}^{L_Z}
|\beta_j|+ \lambda_V \sum_{j=1}^{L_V}
| \gamma_j|,
\]
where $\lambda_Z$ and $\lambda_V$ are pre-determined separate LASSO
penalty parameters for $\beta$ and $\gamma$, respectively, and $w_i$
is an optional sampling weight, which may be used when generalizing
the results obtained from one sample to a target population.\vadjust{\goodbreak}

Our objective function is similar to several existing LASSO variants
but there exist important differences. For example, the elastic net
introduced by \citet{Zou2005} places the same set of covariates under
both a LASSO and ridge constraint to help reduce mis-selections among
correlated covariates. In addition, the group LASSO introduced by
\citet{yuan2006} groups different levels of the same factor together so
that all levels of a factor are selected without sacrificing rotational
invariance. In contrast, the proposed method places separate LASSO
constraints over the qualitatively distinct groups of variables.

\subsection{Estimating heterogeneous treatment effects}

The L2-SVM offers two different means to estimate heterogeneous
treatment effects. First, we can predict the potential outcomes
$Y_i(t)$ directly from the fitted model and estimate the conditional
treatment effect (CTE) as the difference between the predicted outcome
under the treatment status $t$ and that under the control condition,
that is, $\hat
\delta(t;\widetilde{X}_i)=\frac{1}{2}(\widehat{Y}_i(t)-\widehat{Y}_i(0))$.
This quantity utilizes the fact that the L2-SVM is an optimal
classifier [\citet{Lin2002b,Zhang2004}]. Second, we can also estimate
the CATE. To do this, we interpret the L2-SVM as a truncated linear
probability model over a subinterval of $[0,1]$. While it is known
that the SVM does not return explicit probability estimates
[\citet{Lin2002b,Lee2004}], we follow work that transforms the values
$\widehat{W}_i(t)$ to approximate the underlying probability
[\citet{Franc2011,Sollich2002,Platt2000,menon12}]. Specifically, let
$\widehat{W}_i^\ast(t)$ denote the predicted value $\widehat{W}_i(t)$
truncated at positive and negative one. We estimate the CATE as the
difference in truncated values of the predicted outcome variables,
that is, $\hat
\tau(t;\widetilde{X}_i)=\frac{1}{2}(\widehat{W}^\ast_i(t)-\widehat
{W}^\ast_i(0))$.
While this CATE estimate is not precisely a difference in
probabilities, the method provides a useful approximation and returns
sensible results that comport with probabilistic estimates of the
CATE. With an estimated CATE for each covariate profile, the CATE for
any covariate profile can be estimated by simply aggregating these
estimates among corresponding observations.

\subsection{The estimation algorithm}
\label{subsecalgorithm}

Our algorithm proceeds in three steps: the data are rescaled, the
model is fitted for a given value of $(\lambda_Z,\lambda_V)$, and each
fit is evaluated using a generalized cross-validation
statistic.

\textit{Rescaling the covariates}.
LASSO regularization requires rescaling covariates
[\citet{Tibshirani1996}]. Following standard practice, we standardize
all pre-treatment main effects by centering them around the mean and
dividing them by standard deviation. Higher-order terms are
recomputed using these standardized variables. For causal
heterogeneity variables, we do not standardize them when they are
indicator variables representing different treatments. When they
represent the interactions between a treatment indicator variable and
pre-treatment covariates, we interact the (unstandardized) treatment
indicator variable with the standardized pre-treatment
variables.

\textit{Fitting the model}.
The L2-SVM is fitted through a series of iterated LASSO fits, based on
the following two observations. First, we note that for a given
outcome $Y_i^\ast\in\{{\pm} 1\}$, $|1-Y_i^\ast
\widehat{W}_i|_+^2=(Y_i^\ast- \widehat{W}_i)^2\cdot\mathbf{1}\{1
\ge
Y_i^\ast\widehat{W}_i \}$. Thus, the SVM is a least squares problem
on a subset of the data. Second, for a given value of
$\{\lambda_\beta, \lambda_\gamma\}$, rescaling $Z$ and $V$ allows the
objective function to be written as a LASSO problem, with a tuning
parameter of 1, as
\begin{eqnarray*}
\sum_{i=1}^n w_i
\cdot\bigl|Y_i^\ast- \bigl(\mu+ \tilde\beta^\top
\widetilde{Z}_{i} + \tilde\gamma^\top\widetilde{V}_i
\bigr)\bigr|_+^2 + \sum_{j=1}^{L_Z} |
\tilde\beta_j|+ \sum_{j=1}^{L_V} |
\tilde\gamma_j|,
\end{eqnarray*}
where $\tilde\beta=\lambda_\beta\cdot\beta$,
$\tilde\gamma=\lambda_\gamma\cdot\gamma$, $\widetilde{Z}_i = Z_i
/\lambda_\beta$, and $\widetilde{V}_i = V_i /\lambda_\gamma$.

This allows a fitting strategy via the efficient LASSO algorithm of
\citet{Friedman2010}. Specifically, each iteration of our algorithm
consists of fitting a model on the set of ``active'' observations
$\mathcal{A}=\{i\dvtx1 \ge Y_i^\ast\widehat{W}_i\}$ and updating this
set. We now describe the proposed algorithm in greater detail.
First, the values of tuning parameters are selected, $\lambda=\{
\lambda_\beta, \lambda_\gamma\}$. We then select the initial values
of the coefficients and fitted values as follows, $(\tilde\beta^{(0)},
\tilde\gamma^{(0)})=\mathbf{0}$ and $\widetilde{Y}^{(0)}_i=0$ for all
$i$. This places all observations in the initial active set, so that
$\mathcal{A}^{(0)}=\{1,\ldots,n\}$.

Next, for each iteration $k=1,2,\ldots,$ let $\cA^{(k)}= \{i \dvtx1
\ge
Y_i \widehat{W}_i^{(k-1)}\}$ denote the set of active observations, and
$a^{(k)}=\llvert \mathcal{A}^{(k)}\rrvert $ represent the number of observations
in this set. Define $\widetilde{Z}_i^{(k)}$ and $\widetilde{V}_i^{(k)}$
as the centered versions of $\widetilde{Z}_i$ and $\widetilde{V}_i$
(around their respective mean), respectively, using only the
observations in the current active set $\cA^{(k)}$. Similarly, we use
$Y_i^{(k)}$ to denote the centered value of $Y_i^\ast$ using only the
observations in $\cA^{(k)}$. Then, at each iteration $k$, the algorithm
progresses in three steps. We update the LASSO coefficients~as
\begin{eqnarray*}
&&\bigl(\tilde{\beta}^{(k)}, \tilde{\gamma}^{(k)} \bigr)\\
&&\qquad =
\argmin_{(\tilde
{\beta},\tilde{\gamma})} \Biggl\{ \frac{1}{a^{(k)}} \sum
_{i
\in\mathcal{A}^{(k)}} \bigl(Y_i^{(k)}-\tilde {
\beta}^\top\widetilde{Z}_i^{(k)}- \tilde {
\gamma}^\top\widetilde{V}_i^{(k)}
\bigr)^2+ \sum_{j=1}^{L_Z} |\tilde
{\beta}_j|+ \sum_{j=1}^{L_V} |
\tilde {\gamma}_j| \Biggr\}.
\end{eqnarray*}
The fitted value is updated as $\widehat{W}_i^{(k)}= \hat\mu^{(k)}+
{\tilde\beta}^{{(k)}^\top} \widetilde{Z}_i + {\tilde\gamma}^{{(k)}
^\top}
\widetilde{V}_i$ where the intercept is $\hat\mu^{(k)}=
\frac{1}{a^{(k)}} \sum_{i \in\mathcal{A}^{(k)}} (Y_i^\ast
-{\tilde\beta}^{{(k)}^\top} \widetilde{Z}_i -
\tilde{\gamma}^{{(k)}^\top} \widetilde{V}_i)$. These steps are
repeated until $(\beta, \gamma)$ converges.

Work concurrent to ours has developed an algorithm for the
regularization path of the L2-SVM [\citet{yangzou12}]. Future work
can combine our work and that of \citet{yangzou12} in order to
estimate the whole ``regularization surface'' implied by a model with
two constraints.

\textit{Selecting the optimal values of the tuning parameters}. We
choose the optimal values of the tuning parameters, $\{\lambda_Z,
\lambda_V\}$, based on a generalized cross-validation (GCV) statistic\vadjust{\goodbreak}
[\citet{Wahba1990}], so that the model fit is balanced against model
dimensionality. The number of nonzero elements of $|\tilde
\beta|_0+|\tilde\gamma|_0=l$ provides an unbiased degree-of-freedom
estimate for the LASSO [\citet{Zou2007}]. Our GCV statistic is defined
over the observation in the active set $\cA$, as follows:
\[
V(\lambda_Z,\lambda_V) = \frac{1}{n (1- l/a)^2} \sum
_{i\in
\mathcal A} \bigl(Y_i^\ast-
\widehat{W}_i \bigr)^2 = \frac{1}{n (1- l/a)^2} \sum
_{i=1}^n \bigl|1-Y_i^\ast \widehat
W_i\bigr|^2_+,
\]
where the second equality follows from the fact that the observations
outside of $\cA$ does not affect the model fit, that is, $|1-Y_i^\ast
\widehat
W_i|^2_+=0$ for $i \notin\mathcal A$.

Given this GCV statistic, we use an alternating line search to find
the optimal values of the tuning parameters. First, we fix
$\lambda_Z$ at a large value, for example, $e^{10}$, effectively
setting all
causal heterogeneity parameters to zero [\citet{Efron2004}]. Next,
$\lambda_V$ is evaluated along the set of widely spaced grids, for example,
$\log(\lambda_V) \in\{-15,-14,\ldots,10\}$, with the value producing
the smallest GCV statistic selected. Given the current estimate of
$\lambda_V$, $\lambda_Z$ is evaluated along the set of widely spaced
grids, for example, $\log(\lambda_V) \in\{-15,-14,\ldots,10\}$, and the
$\lambda_V$ that produces the smallest GCV statistic is selected. We
alternate in a line search between the two parameters to convergence.
After convergence, the radius is decreased based on these converged
parameter values, and the precision is increased to the desired level,
for example, $10^{-4}$. The final estimates of coefficients are estimated
given the converged value of $(\lambda_Z,\lambda_V)$.

The use of this GCV statistic is reasonable when exploring the degree
to which the treatment effects are heterogeneous. However, if the
goal is to derive the optimal treatment rule, the researchers may wish
to directly target a particular measure of the performance of the
learned policy. For example, following \citet{imaistra11} and
\citet{Qian2011}, we could use the largest average treatment effect as
the statistic (known as ``value statistic'') for cross-validation. In
addition to its computational burden, one practical difficulty of
cross-validation based on the value statistic is that when the total
number of causal heterogeneity variables is large and the sample size
is relatively small as in our applications, we may not have many
observations in the test sample that actually received the same
treatment as the one prescribed by the optimal treatment rule. In
addition, the training sample may have empty cells so that they do not
predict treatment effects for some individuals in the test set. This
makes it difficult to apply this procedure in some situations. The
use of the GCV statistic is also computationally efficient as it
avoids cross-validation.

Comparing computation times across competitors is difficult, as they
can vary dramatically depending on the number of cross-validating
folds (Boosting, LASSO), iterations (Boosting), the number of MCMC
draws and number of trees (BART), and desired precision in estimating
the tuning parameters (LASSO, SVM). The general pattern from our
simulations is that the computational time for the proposed method is
significantly greater than the Bayesian GLM, tree, and the
cross-validated logistic LASSO with a single constraint. It is
comparable to BART and significantly less than cross-validated
boosting.

Finally, in a recent paper, \citet{zhaoetal12} propose a method that
is related to ours. There are several important differences
between the two methods. First, we are primarily interested in feature
selection using LASSO penalties, whereas \citeauthor{zhaoetal12}
focus on prediction using an $l_2$ penalty. While we use a simple
parametric model with a large number of features,
\citeauthor{zhaoetal12} place their method within a nonparametric,
reproducing kernel framework. In kernelizing the covariates, they
achieve better prediction but at the cost of difficulty in
interpreting precisely which features are driving the treatment rule.
Second, we use two separate LASSO penalties for causal heterogeneity
variables and pre-treatment covariates, whereas
\citeauthor{zhaoetal12} do not make this distinction. Third, our
tuning parameter is a GCV statistic, which eliminates the
computational burden of cross-validation as used by
\citeauthor{zhaoetal12}

\section{Empirical applications}
\label{secapplications}

In this section we apply the proposed method to two well-known field
experiments in the social sciences.

\subsection{Selecting the best get-out-the-vote mobilization strategies}
\label{subsecGOTV}

First, we analyze the get-out-the-vote (GOTV) field experiment where
69 mobilization techniques were randomly administered to registered
New Haven voters in the 1998 election [\citet{gerbgree00}]. It is
known in the GOTV literature that there is substantial interference
among voters in the same household [\citet{nick08}]. Thus, to avoid
the problem of possible interference between voters, we focus on
$14\mbox{,}774$ voters in single voter households where $5269$ voters of
them belong to the control group and hence did not receive any of
these GOTV messages. Addressing this interference issue fully
requires an alternative experimental design where the treatment
conditions correspond to different number of voters within the same
household who receive the GOTV message (our method is still applicable
to the data from this experimental design because it can handle
multi-valued treatments). For the purpose of illustration, we also
ignore the implementation problems documented in \citet{imai05} and
analyze the most recent data set.

In our specification, the causal heterogeneity variables $Z_i$
include the binary indicator variables of 192 treatment combinations,
that is, $K_Z = 192$. We include a set of noncausal variables $V_i$,
which consist of the main effect terms of four pre-treatment
covariates (age, member of a majority party, voted in 1996, abstained
in 1996), their two-way interaction terms, and the square of the age
variable, that is, $K_V = 10$.




%
\begin{table}[t]
\tabcolsep=0pt
\caption{Estimated heterogeneous treatment effects for the New Haven
get-out-the-vote experiment}\label{GGEffects}\vspace*{-3pt}
\begin{tabular*}{\textwidth}{@{\extracolsep{\fill}}lcccd{2.2}@{}}
\hline
\multicolumn{4}{@{}c}{\textbf{Get-out-the-vote mobilization strategy}} &
\\[-6pt]
\multicolumn{4}{@{}c}{\hrulefill} & \\
\textbf{Visit}& \textbf{Phone} & \textbf{Mailings} & \textbf{Appeal type} & \multicolumn{1}{c@{}}{\textbf{Average effect}}\\
\hline
 Yes& No & 0 & Any&  3.06 \\
Yes& No & 3& Civic& 2.64 \\
Yes& No & 3& Close& 2.31 \\
 Yes & Civic, Close & 0& Close, Neighbor& 2.04 \\
Yes& No& 1--2& Close& 1.60 \\
 Yes & No & 3 & Civic& 1.50 \\
 Yes & Civic, Close & 1--3 & Civic, Close & 1.46 \\
  Yes& None, Civic/Neighbor, Neighbor & 1--2& Neighbor& 1.46\\
Yes & None, Civic& $0,2$& Civic, Neighbor & 1.46 \\
No& No& 3& Civic& 1.17 \\
No& No& 2& Civic&  1.14 \\
No & Close& $0,1,3$ & Close& 0.81 \\
Yes&  Civic/Blood, Neighbor, Neighbor/Civic & 3& Civic, Neighbor &0.80 \\
 No& Close & 2& Close& 0.74 \\
No& No&  3& Close& 0.70 \\
  Yes& Civic/Blood, Civic & 1--3 & Civic & 0.53 \\
No& No& 3&Neighbor& 0.04\\
 Yes& Civic/Blood& 1--3& Civic& -0.64 \\
 No& Neighbor/Civic& 3& Neighbor&-0.65 \\
No& Civic/Blood& $1,3$& Civic& -0.91 \\
No& Civic/Blood& 2& Civic& -0.99 \\
No& Civic/Blood, Civic& $1,3$& Civic& -2.07 \\
No& No& 2& Close&-2.08 \\
No& Civic/Blood& 2& Civic& -2.14 \\
 No& Civic, Civic/Blood, Neighbor& 1& Civic, Neighbor& -2.60 \\
No& Neighbor & 2& Neighbor& -2.67 \\
No& Neighbor& 3& Neighbor& -3.24 \\
No& Civic& 0--1& Neighbor& -3.49 \\
No& Civic& 2& Neighbor& -3.56 \\
No&Civic& 3& Neighbor& -4.12 \\
\hline
\end{tabular*}
\tabnotetext[]{}{\textit{Note}:
Results are presented in terms of
percentage points
increase or decrease relative to the baseline of no treatment of any
type administered.
Every treatment
combination consists of an assignment to personal visit (Yes or No),
phone call (Donate blood, civic appeal, civic appeal/donate blood,
neighborhood solidarity, civic appeal/neighborhood solidarity, close
election), number of mailings (0--3) and appeal type (neighborhood
solidarity, civic appeal, close election).
Personal visits are uniformly more effective than any
other treatment method, while sending three mailings with a Civic
Responsibility message is the most effective treatment with no
visit. Every mobilization strategy with a phone call, but no personal
visit or mailings,
is estimated with a nonpositive sign.}\vspace*{-3pt}
\end{table}

Of the 192 possible treatment effect combinations, 15 effects are estimated
as nonzero (see Table~\ref{GGCoefs} in \hyperref[app]{Appendix}). As these
coefficients range from main effects to four-way interactions, they
are difficult to\vadjust{\goodbreak} interpret. Instead, we present the estimated
treatment effect for every treatment combination in Table
\ref{GGEffects}. Some of our results are consistent with the prior
analysis. First, canvassing in person is the most effective GOTV
technique. This result can be obtained even from a simple use of the
difference-in-means estimator (estimated 2.69 percentage point with
$t$-statistic of 2.68). Second, every mobilization strategy that
consists of a phone call and no personal visit is estimated with a
nonpositive sign, suggesting that the marginal effect of a phone call
is either zero or slightly negative. Most prominently, phone messages
with a neighborhood appeal or civic appeal decrease
turnout.

The proposed method also yields finer findings than the existing
analyses. For example, since personal canvassing is expensive,
campaigns may be interested in the most effective treatment that does
not include canvassing. We find that three mailings with a civic
responsibility message and no phone calls or personal visits increase
turnout marginally by 1.17 percentage point. This result is similar
to the one independently obtained in another study [\citet{Gerber2008}].
Three mailings with other appeals produce smaller effects (1.17 and
0.04 percentage point increase, resp.).

Finally, the proposed method, upon considering all possible
treatments, produces clear prescriptions. First, in the presence of
canvassing, any additional treatment (phone call, mailing) will lessen
the canvassing's effectiveness. If voters are canvassed, they should
not receive additional treatments. Second, if voters are not
canvassed, they should be targeted with three mailings with a civic
duty appeal. Any other treatment combination will be less
cost-effective, and may even suppress turnout.

\subsection{Identifying workers for whom job training is beneficial}
\label{subseclalonde}

Next, we apply the proposed methodology to the national supported work
(NSW) program. Our analysis focuses upon the subset of these
individuals previously used by other researchers
[\citet{lalo86,dehewahb99}] where the (randomly selected) treatment
and control groups consist of 297 and 425 such workers, respectively.
We consider two aspects of treatment effect heterogeneity. First, we
seek to identify the groups of workers for whom the training program
is beneficial. The program was administered to a heterogeneous
group of workers and, hence, it is of interest to investigate whether
the treatment effect varies as a function of individual
characteristics. Second, we show how to generalize the results based
on this experiment to a target population. Such an analysis is
important for policy makers who wish to use experimental results to
decide whether and how to implement this program in a target
population.

%

For illustration, we generalize the experimental results to the 1978
panel study of income dynamics (PSID), which oversamples low-income
individuals. Within this PSID sample, we focus on 253 workers who had
been unemployed at some point in the previous year to avoid severe
extrapolation. This subsample is labeled PSID-2 in
\citet{dehewahb99}. The differences across the two samples are
substantial. The PSID respondents are on average older ($36$ vs. $24$
years old) and more likely to be married ($74\%$ vs. $16\%$) and have
a college degree\vadjust{\goodbreak} ($50\%$ vs. $22\%$) than NSW participants. The
proportion of blacks in the PSID sample ($40\%$) is much less than in
the NSW sample ($80\%$). In addition, on average, PSID respondents
earned more income (\$7600) than NSW participants (\$3000). All
differences, except for proportion Hispanic, are statistically
significant at the 5\% level.

In our model, the matrix of noncausal variables, $V$, consists of 45
pre-treatment covariates. These include the main effects of age,
years of education, and the log of one plus 1975 earnings, as well as
binary indicators for race, marriage status, college degree, and
whether the individual was unemployed in 1975. We also use square
terms for age and years of education, and every possible two-way
interactions among the pre-treatment covariates are included. The
matrix of causal heterogeneity variables $Z$ includes the binary
treatment and interactions between this treatment variable and each of
the 39 pre-treatment covariates. This yields $K_Z = 45$ and $K_V =
44$.

Using this specification, we first fit the model to the NSW sample to
identify the subpopulations of workers for whom the job training
program is beneficial. Second, we generalize these results to the
PSID sample and estimate the ATE and CATE for these low-income
workers. Unfortunately, sampling weights are not available in the
original data and, hence, for the purpose of illustration, we construct
them by fitting a BART model, using $V$ as predictors
[\citet{Hill2011}]. We then take the inverse estimated probability of
being in the NSW sample as the weights for the proposed method
[\citet{stuaetal11}]. To facilitate comparison between the unweighted
and weighted models, we standardize the weights to have a mean equal
to one. A weight greater than one signifies an observation that is
weighted more highly in the PSID model than in the NSW model. This
allows us to assess the extent to which differences in identified
heterogeneous effects reflect underlying differences in the covariate
distributions between the NSW and PSID samples.

After fitting the model to the unweighted and weighted NSW samples,
the CATE is estimated using the covariate value of each observation,
that is, $\hat\tau(1,X_i)$. The sample average of these CATEs yields an
ATE estimate of $7.61$ and $4.61$ percentage points for the NSW and
PSID samples, respectively. Nonzero coefficients from the fitted
models are shown in Table~\ref{LalondeCoefs} of the \hyperref[app]{Appendix}. As with
the previous example, interpreting high order interactions is
difficult. Thus, we present the groups of workers who are predicted
to experience the ten highest and lowest treatment effects of the job
training program in the NSW (Table~\ref{NSWHighLow}) and PSID sample
(Table~\ref{PSIDHighLow}). The groups most helped, and hurt, by the
treatment were identified by matching the observations in these tables
to the nonzero coefficients.

Across both tables, unemployed Hispanics and highly educated,
low-earning non-Hispanics are predicted to benefit from the program.
Similarly, workers who were older and employed and whites with a high
school degree are identified as those who are negatively affected by
the program. Weights marked with asterisks\vadjust{\goodbreak} in each table indicate
heterogeneous effects that are not identified in the other table. For
example, unemployed Blacks with some college are identified as
beneficiaries only in Table~\ref{NSWHighLow}, while married whites
with no high school degree only appear in Table~\ref{PSIDHighLow}.
This difference is explained by the fact that unemployed blacks with
some college make up 2.7\% of the NSW sample but only 0.4\% of the
PSID sample. Similarly, married whites with no high school degree
make up 15.8\% of the PSID sample and are identified in
Table~\ref{PSIDHighLow}, but only make up 0.1\% of the NSW sample and
are not
identified in Table~\ref{NSWHighLow}. Indeed, when generalizing the
results to a different population, large groups in that population are
more likely to be selected for heterogeneous treatment effects.
Weighting allows us to efficiently estimate heterogeneous treatment
effects in a target population.

%
%
\begin{sidewaystable}
\tablewidth=\textwidth
\caption{Ten highest and lowest treatment effects of job training
program based on the NSW Data}\label{NSWHighLow}
\fontsize{8}{10}{\selectfont{
\begin{tabular*}{\textwidth}{@{\extracolsep{\fill}}ld{3.0}cd{2.0}cccd{5.0}cd{1.3}@{}}
\hline
\multicolumn{1}{@{}l}{\textbf{Groups most helped or hurt}} & \multicolumn{1}{c}{\textbf{Average}} & & & & &
\multicolumn{1}{c}{\textbf{Highschool}} &
\multicolumn{1}{c}{\textbf{Earnings}}
& \multicolumn{1}{c}{\textbf{Unemp.}} & \multicolumn{1}{c@{}}{\textbf{PSID}} \\
\multicolumn{1}{@{}l}{\textbf{by the treatment}} &\multicolumn{1}{c}{\textbf{effect}} &
\multicolumn{1}{c}{\textbf{Age}} & \multicolumn{1}{c}{\textbf{Educ.}}
& \textbf{Race} & \textbf{Married} & \textbf{degree} &
\multicolumn{1}{c}{\textbf{(1975)}}
& \multicolumn{1}{c}{\textbf{(1975)}} & \multicolumn{1}{c@{}}{\textbf{weights}} \\
\hline
\multicolumn{1}{@{}l}{\textit{Positive effects}} \\
 \quad Low education, Non-Hispanic  &53 & 31 & 4 & White & No & No & 10\mbox{,}700 & No & 1.36 \\
\qquad High Earning &  50 & 31 & 4 & Black & No & No & 4020 & No &
0.97*\\[3pt]
  &  40 & 28 & 15 & Black & No & Yes & 0 & Yes & 0.89* \\
\quad Unemployed, Black, &  38 & 30 & 14 & Black & Yes & Yes & 0 & Yes & 1.28*\\
\qquad Some College &  37 & 22 & 16 & Black & No & Yes & 0 & Yes & 0.99*
\\[3pt]
  &  45 & 33 & 5 & Hisp & No & No & 0 & Yes & 0.89\\
&  39 & 50 & 10 & Hisp & No & No & 0 & Yes & 1.28* \\
\quad Unemployed, Hispanic &  37 & 33 & 9 & Hisp & Yes & No & 0 & Yes & 1.13* \\
&  37 & 28 & 11 & Hisp & Yes & No & 0 & Yes & 1.02* \\
&  37 & 32 & 12 & Hisp & Yes & Yes & 0 & Yes & 1.80* \\[3pt]
\multicolumn{1}{@{}l}{\textit{Negative effects}}\\
\quad Older Blacks,  &   -17 & 43 & 10 & Black & No & No & 4130 & No & 1.15 \\
\qquad No HS Degree &  -20 & 50 & 8 & Black & Yes & No & 5630 & No & 4.55
\\[3pt]
 &  -17 & 29 & 12 & White & No & Yes & 12\mbox{,}200 & No & 1.45* \\
 \quad Unmarried Whites, &  -17 & 31 & 13 & White & No & Yes & 5500 & No & 1.56 \\
 \qquad HS Degree &  -19 & 31 & 12 & White & No & Yes & 495 & No & 1.12\\
&  -19 & 31 & 12 & White & No & Yes & 2610 & No & 1.21 \\[3pt]
&  -20 & 36 & 12 & Hisp & No & Yes & 11\mbox{,}500 & No & 1.10* \\
\quad High earning Hispanic &  -21 & 34 & 11 & Hisp & No & No & 4640 & No & 0.89* \\
&  -21 & 27 & 12 & Hisp & No & Yes & 24\mbox{,}300 & No & 0.95* \\
&  -21 & 36 & 11 & Hisp & No & No & 3060 & No & 0.88* \\
\hline
\end{tabular*}}}
\tabnotetext[]{}{\textit{Note}:
Each row represents the estimated
treatment effect given the characteristics of workers. The most
effective treatment rule would target low-education, high income
Non-Hispanics; unemployed blacks with some college, and unemployed
Hispanics. The treatment would be least
effective when administered to older, employed recipients; unmarried whites
with a high school degree but no college; and high earning Hispanics
with no college. The last column represents the PSID weights, which
are the inverse of the estimated probability of being in the NSW
sample, standardized to have mean one. Weights marked with an
asterisk indicate the groups which are not identified as having
highest or lowest treatment effects when generalizing the results to
the PSID sample (see Table~\protect\ref{PSIDHighLow} for those results).}
\end{sidewaystable}

%
%
\begin{sidewaystable}
\tablewidth=\textwidth
\caption{Highest and lowest estimated treatment effects
when generalizing the results to the PSID sample}\label{PSIDHighLow}
\fontsize{8}{10}{\selectfont{\begin{tabular*}{\textwidth}{@{\extracolsep{\fill}}ld{3.0}cd{2.0}cccd{5.0}cd{1.3}@{}}
\hline
\multicolumn{1}{@{}l}{\textbf{Groups most helped or hurt}} & & & & & &
\multicolumn{1}{c}{\textbf{Highschool}} &
\multicolumn{1}{c}{\textbf{Earnings}}
& \multicolumn{1}{c}{\textbf{Unemp.}} & \multicolumn{1}{c@{}}{\textbf{PSID}} \\
\multicolumn{1}{@{}l}{\textbf{by the treatment}} &\multicolumn{1}{c}{\textbf{Effect}} &
\multicolumn{1}{c}{\textbf{Age}} & \multicolumn{1}{c}{\textbf{Educ.}}
& \textbf{Race} & \textbf{Married} & \textbf{degree} &
\multicolumn{1}{c}{\textbf{(1975)}}
& \multicolumn{1}{c}{\textbf{(1975)}} & \multicolumn{1}{c@{}}{\textbf{weights}} \\
\hline
\multicolumn{1}{@{}l}{\textit{Positive effects}} \\
 &86 & 22 & 10 & White & Yes & No & 23\mbox{,}000 & No & 2.14* \\
\quad Married Whites,  &  77 & 20 & 11 & White & Yes & No & 8160 & No & 1.87* \\
\qquad  No HS Degree&  69 & 28 & 10 & White & Yes & No & 6350 & No & 1.64* \\
&  60 & 26 & 8 & White & Yes & No & 36\mbox{,}900 & No & 1.61* \\[3pt]
&  75 & 20 & 12 & White & Yes & Yes & 8640 & No & 5.47* \\
\quad Married, Educated &  50 & 25 & 14 & Black & Yes & Yes & 11\mbox{,}500 & No & 1.82* \\
&  49 & 27 & 13 & White & Yes & Yes & 854 & No & 3.19* \\
&  48 & 24 & 12 & White & Yes & Yes & 24\mbox{,}300 & No & 7.19* \\[3pt]
\quad Unmarried, Uneducated &  83 & 31 & 4 & White & No & No & 10\mbox{,}700 & No & 1.36 \\
 &  49 & 33 & 5 & Hisp & No & No & 0 & Yes & 0.89\* \\[6pt]
\multicolumn{1}{@{}l}{\textit{Negative effects}} \\
&  -41 & 26 & 13 & White & No & Yes & 5400 & No & 1.44* \\
\quad Unmarried Whites, &  -44 & 29 & 12 & White & No & Yes & 12\mbox{,}200 & No & 1.45* \\
\qquad HS Degree&  -44 & 31 & 12 & White & No & Yes & 495 & No & 1.12\; \\
&  -46 & 31 & 12 & White & No & Yes & 2610 & No & 1.21 \\
&  -57 & 31 & 13 & White & No & Yes & 5500 & No & 1.56 \\[3pt]
&  -43 & 43 & 10 & Black & No & No & 4130 & No & 1.15 \\
\quad Older Blacks&  -43 & 42 & 9 & Black & Yes & No & 3060 & No & 1.60* \\
\qquad Employed, No HS Degree&  -47 & 44 & 9 & Black & Yes & No & 10\mbox{,}900 & No & 2.61* \\
&  -60 & 46 & 8 & Black & No & No & 2590 & No & 1.22* \\
&  -100 & 50 & 8 & Black & Yes & No & 5630 & No & 4.55 \\
\hline
\end{tabular*}}}
\tabnotetext[]{}{\textit{Note}: When
administering the treatment in the PSID sample, the most effective
treatment rule would target married whites with no high school
degree; married, educated
non-Hispanics; and unmarried individuals with little education. The treatment will be least effective when
administered to unmarried whites with a high school degree;
high-earning, older blacks with a high school
degree. The last column represents the PSID weights, which are the
inverse of the estimated probability of being in the NSW sample,
standardized to have a mean equal to one. Weights marked with an
asterisk indicate the groups which are not identified as having
highest or lowest treatment effects when fit to the (unweighted) NSW
sample (see Table~\protect\ref{NSWHighLow} for those
results).}
\end{sidewaystable}


\section{Simulation studies}
\label{secsimulations}

In this section we conduct two simulation studies to evaluate the
performance of the proposed method relative to the commonly used
methods: BART (R package \texttt{bayestree}), Bayesian logistic
regression with a noninformative prior (R package \texttt{arm}),
Conditional Inference Trees [\citet{Hothorn2005}; R package \texttt{
party}], Boosting with the number of iterations selected by
cross-validation (as implemented in R package \texttt{ada}), and logistic
regression with a single LASSO constraint and cross-validation on the
``value'' statistic [\citet{Qian2011}; R package \texttt{glmnet}]. The
first set of simulations corresponds to the situation where the goal
is to select a set of the most effective treatments among many
alternatives. The second set considers the case where we wish to
identify a subpopulation of units for which a treatment is most
effective. In both cases, we assume that the treatment $T_i$ is
independent of the observed pre-treatment covariates $X_i$. The
logistic LASSO method is only applied to the second set of simulations
for the reason mentioned in Section~\ref{subsecalgorithm}. Finally,
for each scenario, we examine 4 different sample sizes between $250$
and $5000$ and run $1000$ simulations.

\subsection{Identifying best treatments from a large number of
alternatives}
\label{subsecbest}

We conduct simulations for selecting a set of the best treatments
among a large number of available treatments. We use two settings,
one with the correct model specification and the other with
misspecified models, where un-modeled nonlinear terms are added to the
data generating process. In the simulations with correct model
specification, we have one control condition, 49 distinct treatment
conditions, and~3 pre-treatment covariates. That is, $Z_i$ consists
of 49 treatment indicator variables and $V_i$ is a vector of 3
pre-treatment covariates plus an intercept, that is, $L_Z = 49$ and $L_V
= 4$. Among 49 treatments, 3 of them have substantive effects; the
ATE is approximately equal to $7$, $5$, and $-3$ percentage points,
respectively. The remaining~46 treatment indicator variables have
nonzero but negligible effects, with the average effect sizes ranging
within $\pm1$ percentage point. In contrast, all pre-treatment
covariates are assumed to have substantial predictive
power.\vadjust{\goodbreak}

We independently sample the pre-treatment covariates from a
multivariate normal distribution with mean zero and a randomly
generated covariance matrix. Specifically, an $(L_V\times L_V)$
matrix, $U=[u_{ij}]$, was generated with $u_{ij} \sim\cN(0,1)$ and
the covariance matrix is given by $U^\top U$. The design matrix for
the 49 treatment variables is orthogonal and balanced. The true
values of the coefficients are set as $\beta=\{7.5,3.3,-2,\ldots\}$
and $\gamma=\{50,-30,30\}$, where $\ldots$ denotes 47 remaining
coefficients drawn from a uniform distribution on $[-0.7, 0.7]$.
Finally, the outcome variable $Y_i \in\{-1,1\}$ is sampled according
to the following model; $\Pr(Y_i = 1 \mid Z_i, V_i) = a(Z_i^\top\beta
+ V_i^\top\gamma+ b)$ with $\{a,b\}$ selected such that the
magnitude of the ATEs roughly equals the values specified above.

For the simulations with an incorrectly specified model, we include
unmodeled nonlinear terms based on the pre-treatment covariates in the
data generating process. Specifically, $V_i$ now includes the
interaction term between the first and second pre-treatment covariates
and the square term of the third pre-treatment covariate as well as
the main effect term for each of the three covariates. These
higher-order terms are used to generate the data, but not included as
covariates in fitting any model. As before, the outcome variable is
generated after an affine transformation in order to keep the size of
the ATEs approximately equal to the pre-specified levels given
above.

%
\begin{figure}

\includegraphics[scale=0.91]{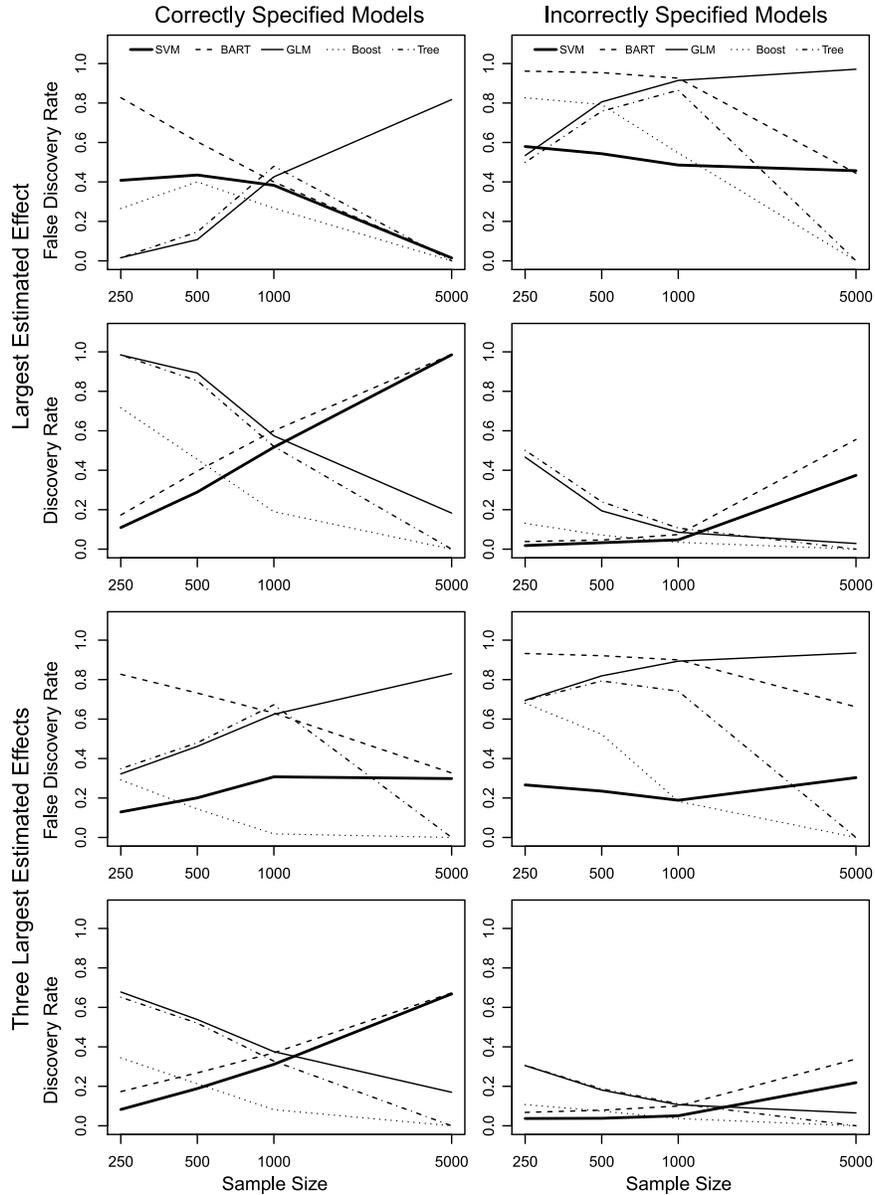}

\caption{False discovery rate (FDR) and discovery rate (DR) for
selecting the best treatments among a large number of available
treatments. Simulation results with correct specification (left
column) and incorrect specification (right column) are shown. The
figure compares the performance of the proposed method (\texttt{SVM};
thick solid lines) to that of BART (\texttt{BART}; dashed lines),
conditional inference trees (\texttt{Tree}; dashed-dotted lines),
Boosting (\texttt{Boost}; dotted lines), and Bayesian logistic
regression with a noninformative prior (\texttt{GLM}; solid lines).
Here, discovery is defined as estimating the largest effect (three
largest effects) as the largest effect (nonzero effects) with the
correct sign. The first and second (third and fourth) rows
present the FDR and DR for the largest effect (three largest
effects), respectively.} \label{figFirstSims}
\end{figure}

Figure~\ref{figFirstSims} summarizes the results in terms of false
discovery rate (FDR) and discovery rate (DR) separately for the
largest and substantive effects. We define discovery as estimating the
largest effect (three largest effects) as the largest effect (nonzero
effects) with the correct sign. Similarly, false discovery occurs
when the largest effect is not correctly discovered \textit{and} at least
one coefficient is estimated to be nonzero. FDR may not equal one
minus DR because the former is based only on the simulations where at
least one coefficient is estimated to be nonzero. The first row
presents FDR for the largest effect whereas the second row presents
its DR. Similarly, the third row plots the FDR for the three largest
effects while the fourth row presents their DR. Note that fewer than
three nonzero effects may be estimated.\looseness=-1

The results show that across simulations the proposed method (\texttt{SVM}; solid lines) has a smaller FDR while its DR is competitive
with other methods. The comparison with BART reveals a key feature of
our method. The proposed method dominates BART in FDR regardless of
model specification. The largest estimated effect from BART
identifies the largest effect slightly more frequently, but at the
cost of a higher FDR. Despite its low FDR, our method maintains a
competitive DR. For many methods, model misspecification increases FDR
and reduces DR. For this reason, we recommend erring in favor of
including too many rather than too few pre-treatment covariates in the
model.

Unlike three of its competitors, Boosting, conditional inference
trees, and Bayesian GLM, the performance of the proposed method
improves as the sample size increases. Boosting\vadjust{\goodbreak} and trees both attain
an FDR and DR of zero as the sample size grows. The tree focuses in
on the largest effects, not identifying any small effects as the
sample size grows. This may be due to the fact that trees do not
converge asymptotically to the true conditional mean function unless
the underlying function is piecewise constant (though they converge to
the minimal risk) [see, e.g., \citet{breietal84}]. The boosting
algorithm uses trees as base learners, which may be leading to the
deteriorating performance in identifying small effects. The
performance of Bayesian GLM also declines with increasing sample size,
because we are not considering uncertainty in the posterior mean
estimates. To address this issue, one must use some $p$-value based
regularization, such as using a $p$-value threshold of $0.10$ (see
Figure \ref{figUsvsBayes} in the next set of simulations for
illustration).

\subsection{Identifying units for which a treatment is beneficial/harmful}

In the second set of simulations, we consider the problem of
identifying groups of units for which a treatment is beneficial (or
harmful). Here, we are interested in identifying interactions between
a treatment and observed pre-treatment covariates. The key difference
between this simulation and the previous one is that in the current
setup causal heterogeneity variables (treatment-covariate
interactions) may be correlated with each other as well as other
noncausal variables. The previous simulation setting assumes that
causal heterogeneity variables (treatment indicators) are independent
of each other and other variables. In this simulation, we also
include a comparison with the logistic regression with a single LASSO
constraint and the maximal ten-fold cross-validation on the ``value''
statistic [\citet{Qian2011}]. This statistic is the expected benefit
from a particular treatment rule [see also \citet{imaistra11}]. Note
that in the previous simulation, due to a
large number of treatments, cross-validation on this statistic is not
feasible.

%
\begin{figure}

\includegraphics{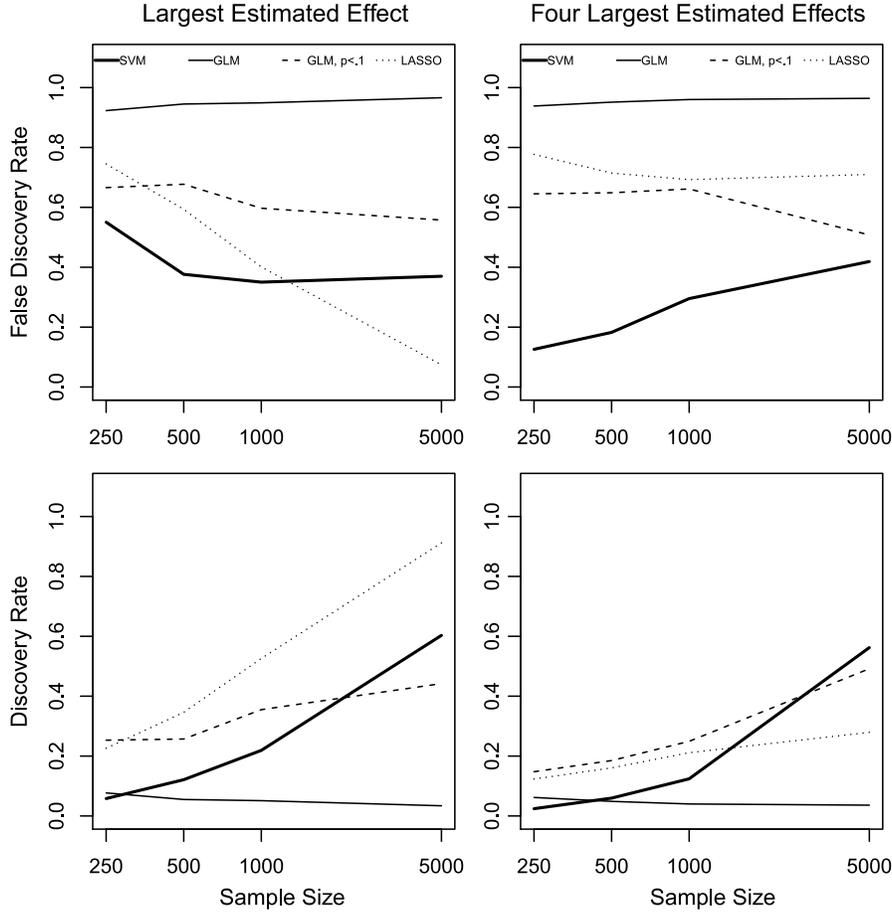}

\caption{False discovery rate (FDR) and discovery rate (DR) for
identifying units for which a treatment is most effective (or
harmful). The figure compares the performance of the proposed
method (\texttt{SVM}; solid lines) with the logistic LASSO (\texttt{LASSO})
and the Bayesian logistic regression based on a
noninformative prior (\texttt{GLM}; dashed and dotted lines). For
Bayesian GLM, we examine the estimates based on posterior means
(dashed lines) and statistical significance ($p$-value less than
$0.1$). Here, discovery is defined as estimating the largest
effect (four largest effects) as the largest effect (nonzero
effects) with the correct sign. The top and bottom plots in the
first (second) column present the FDR and DR for the largest
effect (three largest effects), respectively.} \label{figUsvsBayes}
\end{figure}

In the current simulation, we have a single treatment condition, that is,
$K = 1$, and 20 pre-treatment covariates $X_i$. The pre-treatment
covariates are all based on the multivariate normal distribution with
mean zero and a random variance-covariance matrix as in the previous
simulation study, with five covariates then discretized
using 0.5 as a threshold. Causal heterogeneity variables $Z_i$
consist of 20 treatment-covariate interactions plus the main effect
for the treatment indicator ($L_Z=21$), while $V_i$ is composed of the main
effects for the pre-treatment covariates ($L_V=20$).

Given this setup, we generate the outcome variable $Y_i$ in the same
way as in Section~\ref{subsecbest} according to the linear
probability model. There are 4 pre-treatment covariates that interact
with the treatment in a systematic manner. As before, we apply an
affine transformation so that an observation whose values for these
two covariates are one standard deviation above the mean has the CATE
of roughly~$4$ and $1.7$ percentage points. That is, we set
$\beta=\{-2.7,2.7,-6.7,-6.7,\ldots\}$ and
$\gamma=\{50,-30,30,20,-20,\ldots\}$ where the $\ldots$ denotes
uniform draws from $[-0.7, 0.7]$.

Figure~\ref{figUsvsBayes} compares the FDR and DR for our proposed
method (\texttt{SVM}; solid lines) with those for the logistic LASSO (\texttt{LASSO})
and Bayesian logistic regression (\texttt{GLM}; dotted and dashed lines).
For the Bayesian GLM, we consider two rules: one based on posterior
means of coefficients (dashed lines) and the other selecting
coefficients with $p$-values below $0.1$ (dotted lines). Unlike the
simulations given in Section~\ref{subsecbest}, neither BART,
boosting, nor conditional inference trees provide a simple rule for
variable selection in this setting and hence no results are reported.

The interpretation of these plots is identical to that of the plots in
Figure~\ref{figFirstSims}. In the left column, the top (bottom) plot
presents FDR (DR) for the largest effect, whereas that of the right
column presents FDR (DR) for the four largest effects. When compared
with the Bayesian GLM, the proposed method has a lower FDR for both
largest and four largest estimated effects. The $p$-value
thresholding improves the Bayesian GLM, and yet the proposed method
maintains a much lower FDR and comparable DR. Relative to the LASSO,
the proposed method is not as effective in considering the largest
estimated effect except that it has a lower FDR when the sample size
is small. However, when considering the four largest estimated
effects, the proposed method maintains a lower FDR than the LASSO, and
a comparable DR. This result is consistent with the fact that the
value statistic targets the largest treatment effect while the GCV
statistic corresponds to the overall fit.\looseness=-1

To further evaluate our method, we consider a situation where each
method is applied to a sample and then used to generate a treatment
rule for each individual in another sample. For each method, a
payoff, characterized by the net number of people in the new sample
who are assigned to treatment and are in fact helped by the treatment,
is calculated. To represent a budget constraint faced by most
researchers, we specify the total number of individuals who can
receive the treatment and vary this number within the simulation
study.\looseness=-1

Specifically, after fitting each model to an initial sample, we draw
another simple random sample of 2000 observations from the same data
generating process. Using the result from each method, we calculate
the predicted CATE for each observation of the new sample, $\hat
\tau(1; \widetilde{X}_i)$, and give the treatment to those with
highest predicted CATEs until the number of treated observations
reaches the pre-specified limit. Finally, a payoff of the form
$\mathbf{1}\{\hat\tau(1; \widetilde{X}_i) >0\} \cdot\sgn(\tau(1;
X_i))$ is calculated for all treated observations of the new sample
where $\tau(1; X_i)$ is the true CATE. This produces a payoff of $0.5$
if a treated observation is actually helped by the treatment, $-0.5$ if
the observation is harmed, and $0$ for untreated observations. As a
baseline, we compare each method to the ``oracle'' treatment rule,
$\mathbf{1}\{\tau(1; X_i)>0\}\cdot\sgn(\tau(1; X_i))$, which
administers the treatment only when helpful. We have also considered
an alternative payoff of the form $\mathbf{1}\{\hat\tau(1;
\widetilde{X}_i)>0\} \cdot\tau(1; X_i)$, representing how much
(rather than whether) the treatment helps or harms. The results were
qualitatively similar to those presented here.\looseness=-1

%
%
\begin{table}
\caption{Performance in payoff relative to the oracle}\label{tabpayoff}
\begin{tabular*}{\textwidth}{@{\extracolsep{\fill}}ld{4.0}d{3.0}d{3.0}d{3.0}@{}}
\hline
&\multicolumn{4}{c@{}}{\textbf{Sample size}}\\[-6pt]
&\multicolumn{4}{c@{}}{\hrulefill}\\
\multicolumn{1}{@{}l}{\textbf{Method}}& \multicolumn{1}{c}{\textbf{250}} & \multicolumn{1}{c}{\textbf{500}} & \multicolumn{1}{c}{\textbf{1000}} &
\multicolumn{1}{c@{}}{\textbf{5000}} \\
\hline
\multicolumn{1}{@{}l}{\texttt{SVM}} & -2 & 11 & 22 & 42 \\
\multicolumn{1}{@{}l}{\texttt{BART}} & -19 & -4 & 8 & 21 \\
\multicolumn{1}{@{}l}{\texttt{LASSO}} & -18 & 2 & 15 & 28 \\
\multicolumn{1}{@{}l}{\texttt{GLM}} & -20 & -7 & 7 & 34 \\
\multicolumn{1}{@{}l}{\texttt{Boost}} & -1 & 10 & 18 & 40 \\
\multicolumn{1}{@{}l}{\texttt{Tree}} & 2 & 2 & 2 & 5 \\
\multicolumn{1}{@{}l}{Treat everyone}& -123 & -121 & -121 & -116 \\
\hline
\end{tabular*}
\tabnotetext[]{}{\textit{Note}: The table
presents a payoff for each method as a percentage of the optimal
oracle rule, which is considered as $100\%$. Each method is fit to a
training set, and
the treatment is administered to every person in the validation set
with a predicted improvement. The proposed method (\texttt{SVM})
narrowly dominates
Boosting (\texttt{Boost}), and both the proposed method and Boosting
noticeably outperform all other competitors,
except conditional inference trees (\texttt{Tree}) at sample size 250.
At larger sample sizes
the tree severely underfits. While the proposed method and Boosting
perform similarly by a predictive criterion, Boosting does not return
an interpretable model. \texttt{BART}, \texttt{GLM}, and \texttt{LASSO}
represent the Bayesian Additive Regression Tree, the logistic
regression with a noninformative prior, and the logistic regression
with a single LASSO constraint and cross-validation on the value
statistic. The bottom row presents the outcome if
every observation were treated, indicating that in this simulation
the average treatment effect is negative, but there exists a subgroup
for which the treatment is beneficial.}
\end{table}

The results from the simulation are presented in Table
\ref{tabpayoff}. The table presents a comparison of payoffs, by
method, as a percentage of the optimal oracle rule, which is
considered as $100\%$. The bottom row presents the outcome if every
observation were treated, indicating that in this simulation the
average treatment effect is negative but there exists a subgroup for
which treatment is beneficial.\vadjust{\goodbreak} The proposed method (\texttt{SVM})
narrowly dominates Boosting (\texttt{Boost}), and both the proposed
method and Boosting noticeably outperform all other competitors,
except conditional inference trees (\texttt{Tree}) at sample size 250.
At larger sample sizes, however, the tree severely underfits. While
the proposed method and Boosting perform similarly by a predictive
criterion, Boosting does not return an interpretable model. We also
find that \texttt{SVM} outperforms \texttt{LASSO}, which is
consistent with
the fact that the GCV statistic targets the overall performance while
the value statistic focuses on the largest treatment effect. If
administering the treatment is costless, the proposed method generates
the most beneficial treatment rule among its competitors.

%
\begin{figure}

\includegraphics{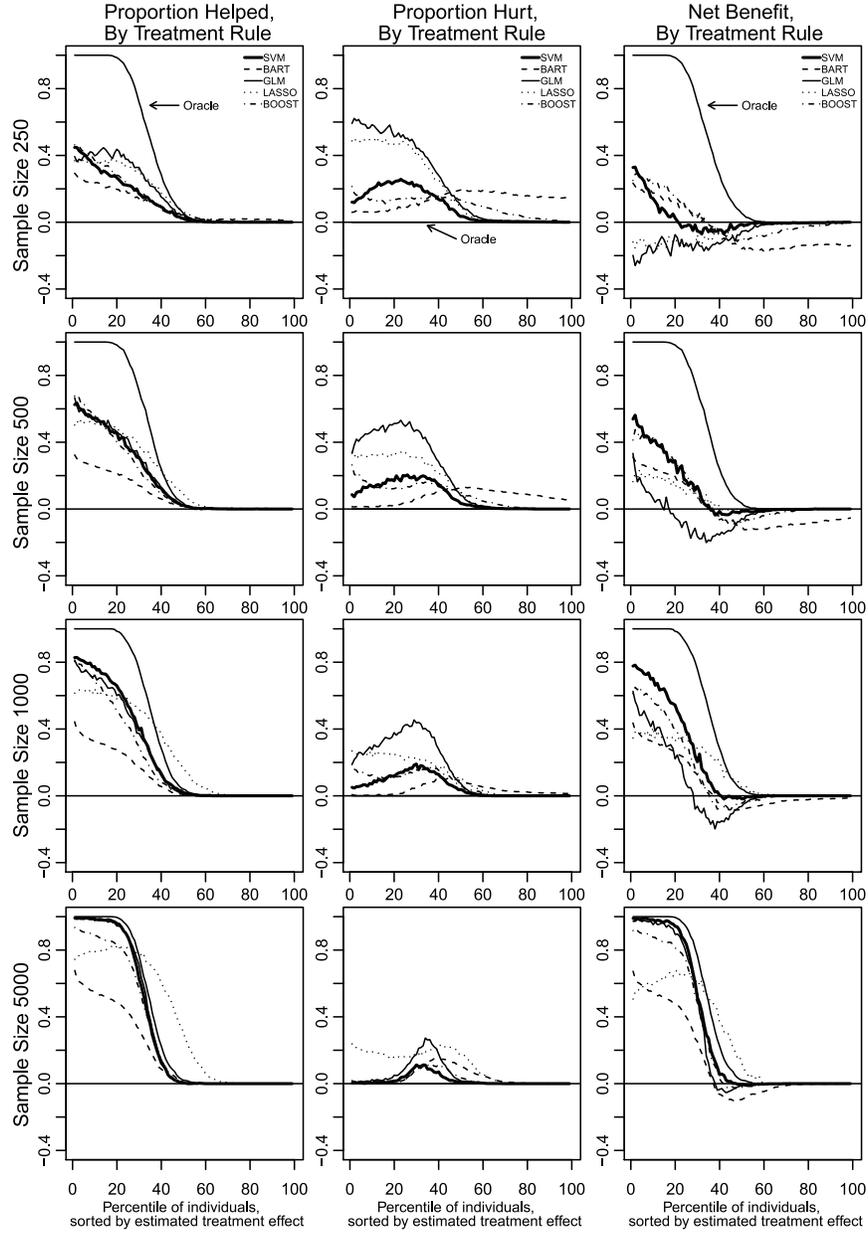}

\caption{The relative performance of individualized treatment rules
derived by each method. The figure presents the proportion of
treated units (based on the treatment rule of each method) who
benefit from the treatment (left column), are harmed by the
treatment (middle column), and the difference between the two
(right column) at each percentile of the total sample who can be
assigned to the treatment. The oracle (solid lines) treats each
observation only when helpful and hence is identical to the
horizontal line at zero in the middle column. The proposed method
(\texttt{SVM}; solid thick lines) makes fewer mistreatments than
other methods, while it is conservative in assigning observations
to the treatment.}
\label{figPosNeg}
\end{figure}

Figure~\ref{figPosNeg} presents the results across methods and sample
sizes in the presence of a budget constraint. The left column shows
the proportion of treated units that actually benefit from the
treatment for each observation considered for the treatment in the
order of predicted CATE (the horizontal axis). The oracle identifies
those who certainly benefit from the treatment and treats them first.
The middle column shows the proportion of treated units that are hurt
by the treatment. Here, the oracle never hurts observations and hence
is represented by the horizontal line at zero. The right column
presents the net benefit by treatment rule, which can be calculated as
the difference between the positive (left column) and negative (middle
column) effects. Each row presents a different sample size to which
each method is applied.

The figure shows that when the sample size is small, the proposed
method assigns fewer observations a harmful treatment, relative to its
competitors. For moderate and large sample sizes, the proposed method
dominates its competitors in both identifying a group that would
benefit from the treatment and avoiding treating those who would be
hurt. This can be seen from the plots in the middle column where the
result based on the proposed method (\texttt{SVM}; solid thick lines)
stays close to the horizontal zero line when compared to other
methods. Similarly, in the right column, the results based on the
proposed method stay above other methods. When these lines go below
zero, it implies that a majority of treated observations would be
harmed by the treatment. The disadvantage of the proposed method is
its conservativeness. This can be seen in the left column where at
the beginning of the percentile the solid thick line is below its
competitors for small sample sizes. This difference vanishes as the
sample size increases, with the proposed method outperforming its
competitors. In sum, when used to predict a treatment rule for
out-of-sample observations, the proposed method makes fewer harmful
prescriptions and often yields a larger net benefit than its
competitors.

\section{Concluding remarks}
\label{secconclusion}

Estimation of heterogeneous treatment effects plays an essential role
in scientific research and policy making. In particular, researchers
often wish to select the most efficacious treatments from a large
number of possible treatments and to identify individuals who benefit
most (or are harmed) by treatments. Estimation of treatment effect
heterogeneity is also important when generalizing experimental results
to a target population of interest.

The key insight of this paper is to formulate the identification of
heterogeneous treatment effects as a variable selection problem.
Within this framework, we develop a Support Vector Machine with two
separate sparsity constraints, one for a set of treatment effect
heterogeneity parameters of interest and the other for observed
pre-treatment effect parameters. This setup addresses the fact that
in many applications, pre-treatment covariates are much more powerful
predictors than treatment variables of interest or their interactions
with covariates. In addition, unlike the existing techniques such as
Boosting and BART, the proposed method yields a parsimonious model
that is easy to interpret. Our simulation studies show that the
proposed method has low false discovery rates while maintaining
competitive discovery rates. The simulation study also shows that the
use of our GCV statistic is appropriate when exploring the treatment
effect heterogeneity rather than identifying the single optimal
treatment rule.\vadjust{\goodbreak}

%
%
\begin{table}[b]
\caption{Nonzero coefficient estimates for the New Haven
get-out-the-vote experiment}\label{GGCoefs}
\fontsize{8.1}{10.1}{\selectfont{\begin{tabular*}{\textwidth}{@{\extracolsep{\fill}}lcccd{2.2}@{}}
\hline
\multicolumn{4}{c}{\textbf{Treatment interactions}} &  \\[-6pt]
\multicolumn{4}{@{}l}{\hrulefill} &  \\
\textbf{Visit}& \textbf{Phone} & \textbf{Mailings} &\textbf{Message}& \multicolumn{1}{c@{}}{\textbf{Coefficient}}\\
\hline
 Yes &Any & 0--3 & Any&1.50 \\
No& No & 2& Neighborhood & 1.25 \\
Yes& No & 0 & Any&1.04 \\
No & Close & 0 & Close&  0.84 \\
Any& No & 3 & Any& 0.72 \\
Yes& Any & 0 & Any& 0.59 \\
Any & No & 3 & Civic & 0.49 \\
Yes& No&0--3& Close&  0.15 \\
No & Any& 2 & Any& -0.08 \\
Any& Any& 3& Civic&-0.68 \\
Any& Civic & 0--3& Civic& -0.95 \\
No & No & 2& Close & -2.09 \\
Any& Civic/Blood& 0--3& Civic & -2.16 \\
No & Neighbor &0--3& Neighbor & -2.72 \\
No & Civic& 0--3& Neighbor&-3.67 \\
\hline
\end{tabular*}}}
\tabnotetext[]{}{\textit{Note}: Coefficients are presented on the
percentage point scale. 13 of the 193 causal heterogeneity
parameters were estimated as nonzero ($6.7\%$). The largest positive
coefficients correspond with a personal visit, while the largest
negative effects correspond with receiving a civic or neighborhood
solidarity appeal via phone. These coefficients generate the
predicted treatment effects in Table
\protect\ref{GGEffects}.}
\end{table}

%
%
\begin{table}
\caption{Nonzero coefficient estimates for the job training program
data}\label{LalondeCoefs}
%
\begin{tabular*}{\textwidth}{@{\extracolsep{\fill}}ld{3.2}d{3.2}@{}}
\hline
& \multicolumn{1}{c}{\textbf{NSW}} & \multicolumn{1}{c@{}}{\textbf{PSID}} \\
\hline
Treatment intercept& 6.92 & 6.67 \\
\emph{Main effects} \quad \quad\\
\quad  Age & 0.00 & -0.83 \\
\quad  Married & 1.32 & 3.39 \\
\quad White & 0.00 & 0.10 \\
\emph{Squared terms} \quad \quad\\
\quad  Age$^2$ & -0.03 & -0.09 \\
\quad  Education$^2$  & 0.89 & 0.86 \\
\emph{Interaction terms} \quad \quad\\
\quad  No HS degree, Unemployed in 1975 & -1.06 & 0.00 \\
\quad  White, Married & 0.00 & 26.16 \\
\quad  White, No HS degree & 25.35 & 30.65 \\
\quad  Hispanic, Logged 1975 earnings & -49.36 & -62.15 \\
\quad  Black, Logged 1975 earnings & 8.29 & 0.00 \\
\quad  White, Education& 0.00 & -1.41 \\
\quad  Married, Education & 4.90 & 12.11 \\
\quad Married, Logged 1975 earnings & 0.00 & 5.72 \\
\quad  Education, Unemployed in 1975 & 7.52 & 9.59 \\
\quad Age, Education & 0.00 & -0.47 \\
\quad Age, Black  & -0.56 & 0.00 \\
\quad  Age, Hispanic & 0.00 & 0.34 \\
\quad  Age, Unemployed in 1975 & 3.30 & 4.79 \\
\hline
\end{tabular*}
\tabnotetext[]{}{\textit{Note}: The table presents the estimates from the model fitted
to the NSW Data without (left column) and with the PSID weights (right
column). Coefficients are
rescaled to the percentage point scale. The first row contains the
estimated intercept, which corresponds to the estimated CATE for an
observation with characteristics set at the
mean of all pre-treatment covariates.}
\end{table}

A number of extensions of the method developed in this paper are
possible. For example, we can accommodate other types of outcome
variables by considering different loss functions. Instead of the GCV
statistic we use, alternative criteria such as AIC or BIC statistics
as well as more targeted quantities such as the average treatment
effect for the target population can be employed. While we use LASSO
constraints, researchers may prefer alternative penalty functions such
as the SCAD or adaptive LASSO penalty. Furthermore, although not
directly examined in this paper, the proposed method can be extended
to the situation where the goal is to choose the best treatment for
each individual from multiple alternative treatments. Finally, it is
of interest to consider how the proposed method can be applied to
observational data [e.g., see \citet{zhanetal12} who develop a doubly robust
estimator for optimal treatment regimes] and
longitudinal data settings where the derivation of optimal dynamic
treatment regimes is a frequent goal
[e.g., \citet{murp03,zhaoetal11}]. The development of such
methods helps applied researchers avoid the use of ad hoc subgroup
analysis and identify treatment effect heterogeneity in a
statistically principled manner.

%
\begin{appendix}\label{app}

\section*{Appendix: Estimated nonzero coefficients for empirical
applications}

\end{appendix}

\newpage
\section*{Acknolwedgments} An earlier version of this paper was
circulated under the title of ``Identifying Treatment Effect
Heterogeneity through Optimal Classification and Variable
Selection'' and received the Tom Ten Have Memorial Award at the
2011 Atlantic Causal Inference Conference. We thank Charles Elkan, Jake
Bowers, Kentaro Fukumoto, Holger Kern, Michael Rosenbaum, and
Sherry Zaks for useful comments. The Editor, Associate
Editor, and two anonymous reviewers provided useful advice.

%


\printaddresses

\end{document}